\documentclass[useAMS]{mn2e}
\usepackage{epsfig}
\usepackage{graphicx}

\title[]{A Multi-wavelength study of the M dwarf binary YY Geminorum}

\author[]{C.J. Butler$^{1}$ \thanks{E-mail : cjb@arm.ac.uk}, 
N. Erkan$^{2}$,
E. Budding$^{3}$, 
J.G. Doyle$^{1}$, 
B. Foing$^{4}$,
G.E. Bromage$^{5}$,
\newauthor B.J. Kellett$^{6}$,
M. Frueh$^{7}$,
J. Huovelin$^{8}$,
A. Brown$^{9}$,
J.E. Neff$^{10}$\\
$^{1}$Armagh Observatory, College Hill, Armagh, BT61 9DG, N. Ireland, UK\\
$^{2}$Physics Dept., \c{C}anakkale Onsekiz Mart University, \c{C}anakkale, Turkey\\
$^{3}$Carter Observatory, School of Chemical and Physical Sciences, Victoria University, Wellington, New Zealand\\
$^{4}$ESA, Postbus 299, 2200, AG Nordwijk, The Netherlands\\
$^{5}$Jeremiah Horrocks Institute, University of Central Lancashire, Preston, UK\\
$^{6}$Space Science and Technology Department, STFC Rutherford Appleton Laboratory, Oxon, UK\\
$^{7}$McDonald Observatory, 3640 Dark Sky Drive, Texas, USA\\
$^{8}$Division of Geophysics and Astronomy, Department of Physics, University of Helsinki, Finland\\
$^{9}$Center for Astrophysics and Space Astronomy, University of Colorado, Boulder, CO, USA\\
$^{10}$Department of Physics and Astronomy, College of Charleston, Charleston, SC, USA
}

\begin{document}

\maketitle

\begin{abstract}

We review the results of the 1988 multi-wavelength campaign on the late-type 
eclipsing binary YY Geminorum. Observations include: broad-band optical and
near infra-red photometry, simultaneous optical and ultraviolet (IUE) spectroscopy,
X-ray (Ginga) and radio (VLA) data. From models fitted to the optical light curves, 
fundamental physical parameters have been determined together with evidence for 
transient maculations (spots) located near quadrature
longitudes and intermediate latitudes.

Eclipses were observed at optical, ultraviolet and radio wavelengths. 
Significant drops in 6cm radio emission near the phases of both primary and 
secondary eclipse indicate relatively compact radio emitting volumes that may
lie between the binary components. IUE observations during secondary 
eclipse are indicative of a uniform chromosphere saturated with MgII plage-type 
emission and an extended volume of Ly$\alpha$ emission. 

Profile fitting of high-dispersion H$\alpha$ spectra confirms the chromospheric 
saturation and indicates significant H$\alpha$ opacity to heights 
of a few percent of the photospheric radius. There is evidence for an enhanced 
H$\alpha$ emission region visible near phase 0.25-0.35 which may be associated with a large spot on the 
primary and with two small optical flares which were also observed at other wavelengths: one in microwave
radiation and the other in X-rays. For both flares, L$_{X}$/L$_{opt}$ is consistent 
with energy release in closed magnetic structures.

\end{abstract}


\section{Introduction}

YY Geminorum, (BD +32 1582, SAO 60199, Gliese 278c), is a short period (19.54
hours) eclipsing binary with two almost identical  dM1e (flare star)
components.  The close binary is a subsystem of  the nearby Castor multiple star
(YY Gem = Castor C), at a distance of $\sim$14.9 pc. The binary nature was
discovered in 1916 (Adams \& Joy, 1917) and the first spectroscopic orbits were
given  by Joy \& Sanford (1926). As the brightest known eclipsing binary of the
dMe type, YY Gem is an  important fundamental standard for defining the low-mass
Main Sequence mass-luminosity and mass-radius relationships   (Torres \& Ribas,
2002). However, it was clear already from Kron's (1952) pioneer study that there
are significant  surface inhomogeneities  (starspots) affecting the observed
brightness of both components, likely to complicate data analysis.  YY Gem was
the first star,  after the Sun, in which such maculation effects were
demonstrated. Before we can accurately define the intrinsic  luminosities of
such  stars we need to clarify the scale of these effects.  This is also
significant for comparing the photometric  parallax with direct  measurements,
such as that from HIPPARCOS (Budding et al., 2005). 

The system was reviewed by Torres \& Ribas (2002) and Qian et al.\ (2002), the
latter concentrating mainly on  apparent variations  of the orbital period.
Torres \& Ribas (2002) gave revised values for the mean mass and radius of the
very similar  components as  (solar units) $M$ = 0.5992$\pm$0.0047, $R$ =
0.6191$\pm$0.0057, with mean effective temperature  $T$ = 3820$\pm$100 K, as
well as an improved parallax for the system of 66.90$\pm$0.63 mas. From such
results, Torres \& Ribas argued that there had been a tendency to adopt
systematically erroneous parameters for  dwarf stars comparable to YY Gem, with
wider implications for low-mass stars in general.

Determination of the precise structure of these stars, in view of the absence of
definitive information on their intrinsic, spot-free, luminosities,  is still
rather  an open question. Torres \& Ribas (2002) and Qian et al.\ (2002) revised
the work of Chabrier \& Baraffe (1995),  giving radiative  core radii of about
70\%, leaving the outer 30\% to account for the convective zone.  The strong
subsurface  convective motions,  give rise to large-scale magnetic fields that
produce large starspots (cf.\ Bopp \& Evans, 1973). Moffet first reported  large flare
events,  and, in subsequent studies, YY Gem has been shown to be very active 
(Moffet and Barnes, 1979; Lacy et al., 1976; Doyle \& Butler,  1985; Doyle \& Mathioudakis, 
1990; Doyle et al. 1990).

Doyle et al (1990) have previously  described
photometric observations of repetitive, apparently periodic, flares on YY Gem
which were observed during this  programme. More recently, Gao et al.\ (2008)
modelled such periodicity effects on the basis  of magnetic reconnection 
between loops on the two stars generating interbinary flares. Fast
magneto-acoustic waves in plasma trapped in the  space between the two
components  are thought to modulate the magnetic reconnection, producing a
periodic behaviour of the flaring rate. Doyle et al.\ (1990) had previously
suggested filament oscillations. Several authors, (see Vrsnak et al. 2007) have 
subsequently reported solar 
filament oscillations of similar duration to those suggested on YY Gem.

Multi-wavelength observations of flare activity on YY Gem were initiated by
Jackson, Kundu \& White (1989) using radio  data from the VLA (see also, Gary,
1986). Stelzer et al.\ (2002)  used  the Chandra and XMM-Newton satellites in
simultaneous observations  of the X-ray  spectrum, while Saar \& Bookbinder
(2003) carried out far ultraviolet observations.  Impulsive UV and X-ray
phenomena,  taken to  be essentially flare-like, were shown to be orders of
magnitude stronger than those occurring on the Sun (Haisch et al., 1990). 
Tsikoudi \&  Kellett (2000), reviewing X-ray and UV observations of the Castor
system, reported a large (EXOSAT) flare event with  total X-ray  emission
estimated as $\sim$7$\pm 1 \times 10^{33}$ ergs.  Their  comparison of X-ray
and bolometric heating  rates pointed to strong magnetic  activity within hot
coronal components.

\begin{figure*} 
\centering 
\vspace{9cm}
\vspace*{-8cm}

\includegraphics[width=13cm,clip,angle=0]{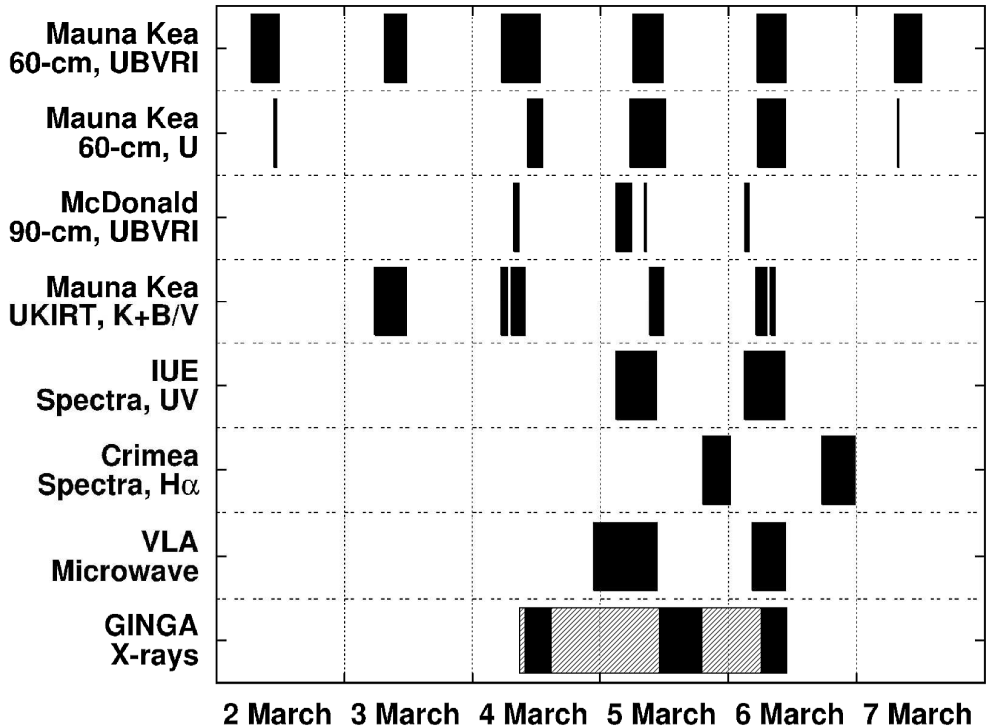}
 
\vspace*{1cm}
\caption{Time-line of
various facilities used in the multiwavelength campaign of 1988} 

\end{figure*}

\begin{table}
\begin{center}
\caption{Multiwavelength Observations of YY Gem, March 1988}

   \begin{tabular}{lccc}

Institute & Observer & Facility & Range\\
\hline
ISAS-Tokyo & Bromage & GINGA & ME X-rays  \\
VILSPA-ESA& Foing & IUE & UV \\
Mauna Kea & Butler & UKIRT  & IR  \\
Mauna Kea & Doyle/Butler & 0.6m & UBVRI \\
McDonald Obs.&  Frueh  & 0.9m MCD & UBVRI  \\
JILA Boulder& Brown & VLA & 5 \& 1.4 GHz  \\
Crimea Obs.&  Tuominen & 2.6m Shajn & UBVRI + H~${\alpha}$ \\
\hline \\
\end{tabular}
\end{center}
\end{table}

In this article, we concentrate on the multiwavelength campaign   initiated from
the Armagh Observatory in 1988 (Butler, 1988). Our general aim is to bring
together results of  some work, previously reported (e.g. Doyle et al. 1990, Butler, et al. 1994, 1995,
Budding et al. 1996, Tuominen et al. 1989) with contemporaneous satellite and radio observations 
thereby allowing an overview of
the campaign. One specific intention concerns the various light curves and their
analyses in terms of standard eclipsing binary models that include photospheric
inhomogeneities.  In addition, we present hitherto unpublished, ultraviolet
(IUE), radio (VLA) and X-ray (Ginga) data, which should be relevant to
subsequent studies.  A number of optical flares were observed but only two of
these were seen at other wavelengths, one in X-rays by Ginga and the other in
the microwave region by the VLA. 

\section[]{The 1988 multi-wavelength campaign on YY Geminorum} In late February
to early March 1988, YY Gem was the object of a coordinated multiwavelength
campaign to observe the star simultaneously in radio, near infra-red, X-rays, UV
and optical radiation (Butler, 1988). The principal objectives of this programme
were: (i) To provide multicolour photometry of the light curve in order to
establish (a) the distribution of surface inhomogeneities (starspots), and (b)
the temperature difference of these inhomogeneous regions from the normal
photosphere. (ii) To provide high time-resolution photometry in V and K during
the eclipses in order to check on possible surface inhomogeneities by  `eclipse
imaging'  --- i.e.\ examining any small  disturbances observed in the light
curve during eclipses.  (iii) To use optical spectroscopy, X-ray and radio
monitoring to probe the outer atmospheres of the components and assess any
topographical connection between photospheric spots and bright chromospheric or
coronal regions. (iv)  To monitor flares on YY Gem in as many separate wavebands
as possible in order to check their energy distribution  and constrain 
models.  

The programme involved the facilities and observations given in Table 1. Several
other organisations offered support to the campaign, but unfortunately a 
number of these were unable to provide useful data due to poor observing
conditions.  In Figure 1, we show the overlap between observing facilities that
were successful in obtaining data. Seven major facilities provided the most
relevant data and six of these were operative on Mar 5 and 6, with a few hours
of overlap on those two days; and to a lesser extent on Mar 4.  

\begin{figure*}
\centering
\includegraphics[width=13cm,clip,angle=0]{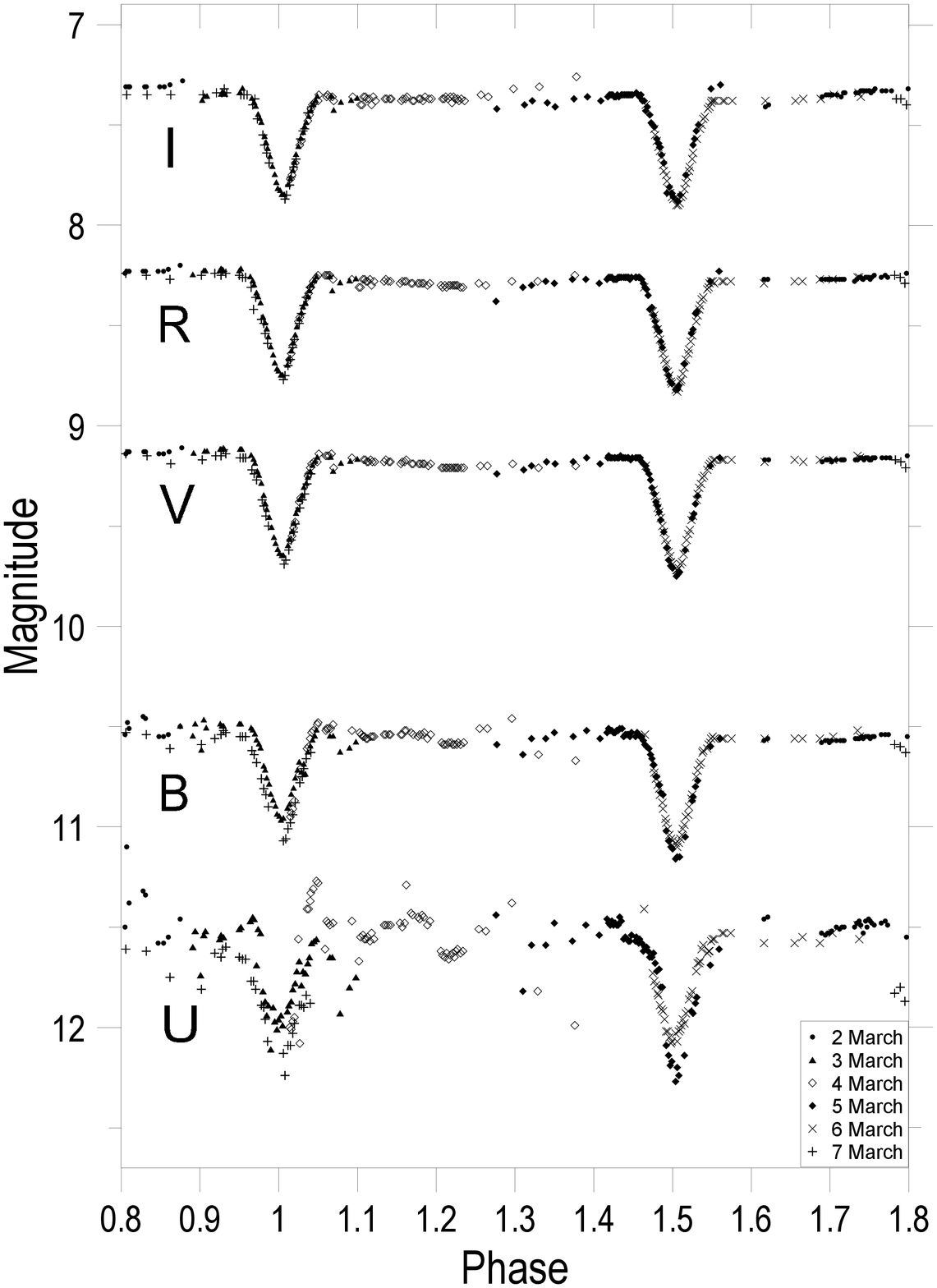}

\caption{Hawaii 0.6 m UBVRI light curves of YY Gem.}

\end{figure*}

\begin{figure*}
\centering
\includegraphics[width=14cm,clip,angle=0]{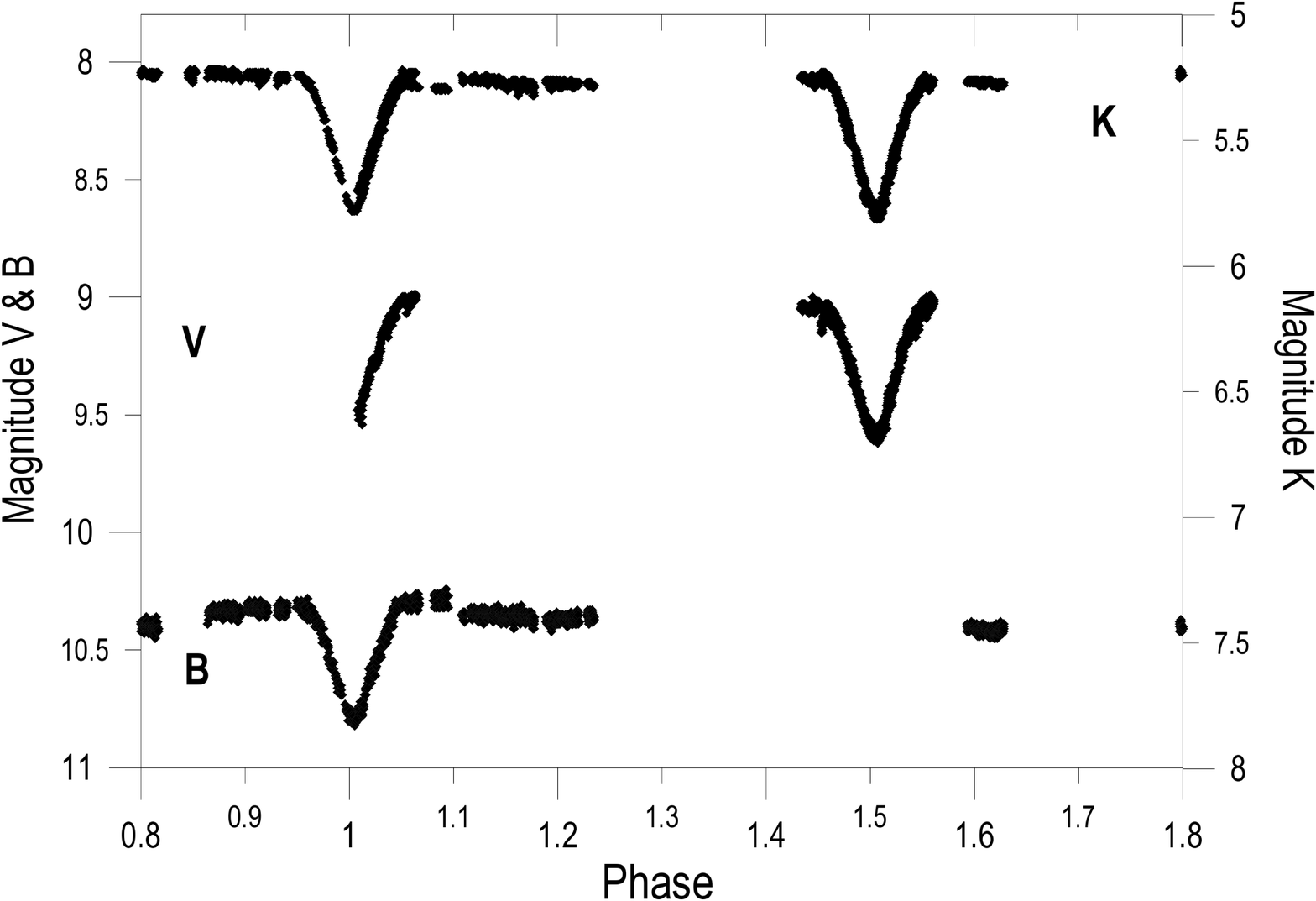}

\caption{UKIRT-BVK light curves of YY Gem.}

\end{figure*}

\section{UBVRIK photometry}
\subsection{Photometric Techniques}
To achieve the photometric aims we required broad-band photometry covering as
much of the optical and infra-red regions as possible. We therefore operated two
telescopes simultaneously: the University of Hawaii 0.6$^m$ telescope on Mauna
Kea and the neighbouring 3.8$^m$ United Kingdom Infra-Red Telescope (UKIRT). 
Some additional observations were contributed by
Marion Frueh of McDonald Observatory, Texas.

All observers were alerted to a particular problem associated with photometry of YY Gem, 
namely that the close proximity (separation $\sim$71 arcsec) to
YY Gem of the bright star Castor (A2 type, V $\sim$1.6) 
makes it difficult to obtain repeatable and
consistent sky background measurements,  particularly in the U and B bands,
where YY Gem is weak and Castor bright. Kron (1952) commented that, in the
vicinity of YY Gem, 30\% of the monitored  blue light originated with Castor and
only 70\% with YY Gem itself (Budding \& Kitamura 1974). For this campaign, in order to
reduce the errors associated with scattered light, observers 
were requested to take the mean of two adjacent sky areas, one to the east and another
to the west of YY Gem. Frequent reference to three nearby comparison stars: BD
32\degr 1577, BD 31\degr 1611 and BD 31\degr 1627, together with 
standard transformation equations
and mean extinction coefficients allowed a photometric accuracy $\sim$ 0.01 magnitudes
to be achieved. Lists of the standards used, the colour equations
derived and the reduced photometric observations are given in the supplementary electronic tables
 (http://star.arm.ac.uk/preprints/2014/654/).

\subsection{UBVRIK photometry from the 0.6m telescope on Mauna Kea}

The UBVRI photometry, from the 0.6$^m$ telescope and Tinsley Photometer, was
standardised to the Johnson UBV and Cape/Kron RI systems  using equatorial and
southern hemisphere standards from Cousins (1980, 1984). The following mean
extinction coefficients were adopted: $\kappa_U = 0.22$,  $\kappa_B = 0.16$,
$\kappa_V = 0.12$, $\kappa_R = 0.10$ and  $\kappa_I = 0.07$. Due to the manual
operation of the Tinsley photoelectric photometer, time-resolution for a single complete 
UBVRI set of measurements was restricted to several minutes. This was
satisfactory for the slower variations associated with eclipse effects and the
rotational modulation of spots, but unsuitable for flare monitoring. Therefore,
two modes of observation were used on this telescope: (1) UBVRI photometry, with
low  time-resolution ($\Delta{\it t}\sim$ 2$^{\rm m}$) during eclipses and
approximately once per hour at other phases, and  (2) continuous U-band
monitoring at (mainly) out-of-eclipse phases. Some of the latter data was
reported on by Doyle et  al.\ (1990).

Because the 0.6$^m$
telescope was set manually it seems likely that  small errors in positioning of
the background comparison region could be responsible for some of the scatter in
the U and B light curves which increases at shorter wavelengths. However, small,
unrecognised, flares would also contribute to the scatter. In Figure 2, we show
the UBVRI light curves for YY Gem from the combined data obtained on 2-7 March
1988 with the 0.6$^m$ telescope. 

\subsection{BVK photometry with UKIRT on Mauna Kea}

The United Kingdom Infrared Telescope (UKIRT) was scheduled to observe on four 
half-nights, during which two primary and two secondary eclipses occurred.
Continuous monitoring in the K-band simultaneously with V or B was made possible
with a dichroic filter and VISPHOT, a photoelectric photometer set up to
monitor the reflected optical beam. A nodding secondary mirror provided rapid
and repeatable background correction. As spot modulation effects are relatively
more prominent in V and flare effects in B,  it was decided to monitor in K and
V during eclipses and in K and B at out-of-eclipse phases. Useful coverage of
the  out-of-eclipse phases by  UKIRT turned out to be quite limited, however.
The auto-guider was not functional at this time, resulting in occasional guiding
errors. We used the mean atmospheric extinction coefficients given above in 
carrying out the differential reductions (see also Krisciunas et al., 1987).
A selection of standards suitable for both optical and infra-red photometry
was made for the determination of the colour equations. One of these (Gleise
699, Barnard's Star) was believed to be in the declining stage of a flare during
observation on 4 March 1988. Further details are given in the supplementary electronic tables.

In Figure 2, discrepancies can be seen at some phases in the B light curves,
but there is generally good agreement  in V. This is consistent with the greater
influence of background irregularities and small flares at shorter wavelengths.
In Figure 3 we show the UKIRT B, V and K observations. The two broadband
photometric  data-sets (Hawaii 0.6$^m$ and UKIRT) are comparable over common
phase intervals, although the less-scattered  UKIRT data has poorer phase
coverage.

The cool-spot hypothesis receives support from the smaller amplitude of the
out-of-eclipse variation at the longer  wavelengths. This is quite noticeable in
the UKIRT K-band, but less so in the 0.6$^m$ I-band data. 

\subsection{UBVRI photometry from McDonald Observatory} 

In order to increase the
probability of obtaining simultaneous optical photometry with radio, X-ray or
ultraviolet observations of flares, YY Gem was placed on the schedule for the
0.91$^{m}$ telescope at McDonald Observatory, Texas on 4, 5 and 6th March 1988. The
photoelectric McD photometer was equipped with a cooled EMI 9658A photomultiplier.
With sequential exposures through U,B,V,R and I filters of the Johnson system, a
time resolution in each waveband of approximately 20 seconds was obtained.
Unfortunately, a computer crash caused the loss of the electronically recorded
data and it was necessary to manually type in the raw photon counts from the
printed output (as had also been necessary for the UKIRT data for similar
reasons). Transformations to the Johnson UBVRI system relied on observations of
17 stars listed by Moffat \& Barnes (1979) and the three local standards listed in Section
3.1 (Butler,
1988). The following mean extinction coefficients were employed: $\kappa_U =
0.57$,  $\kappa_B = 0.29$, $\kappa_V = 0.17$, $\kappa_R = 0.12$ and  $\kappa_I =
0.09$. As at Mauna Kea we had transformed to the Kron/Cousins R,I system,
rather than Johnson's, the McDonald data was further transformed to the
Kron/Cousins system using equations formulated by Bessell (1979). The UBVRI
observations of YY Gem on the three nights 4/5/6 March 1988 are listed in the
supplementary electronic tables. Though no very large optical flares were
recorded at McDonald during this campaign, a flare of approximately 0.6
magnitudes in U was observed simultaneously with a substantial increase in the
6-cm microwave flux recorded by the VLA.

\section{Modelling the Mauna Kea V Light Curves} 

The idea of large-scale inhomogeneities in the local surface brightness of stars 
is not new, and, after a period of dormancy, was revived
 in the mid-twentieth century, particularly after discussion
 of  possible causes of stellar brightness variation by the careful photometrist 
 G.\ Kron (1947, 1950, 1952).  Subsequently, evidence has accumulated from across 
 the electromagnetic spectrum of magneto-dynamic activity effects on cool stars 
  of a few orders of magnitude greater scale than that known for the Sun. 
  These effects include large areas of the photosphere (spots) with cooler than 
  average temperature.

This subject formed the theme of IAU Symposium  176 (Strassmeier \& Linsky, 1996), 
and was reviewed in Chapter 10 of Budding \& Demircan (2007), which outlines the 
methodology pursued in this paper.   Of course, the use of uniform circular areas 
to model maculation effects is a physical  over-simplification, but it is a 
computational device that allows 
an easily formulated fitting function to match the data to the available 
photometric resolution.  Even with the highest S/N data currently available, 
a macula less than about  5 deg in angular mean radius produces light curve losses only at 
the milli-magnitude level.  Whether a given maculation region's shape is circular, 
or of uniform intensity is unfortunately not recoverable. Other indications on surface structure however, 
such as come from the more detailed Zeeman Doppler Imaging techniques for example 
(Donati et al, 2003), tend to support somewhat simple and uniform structures 
to maculae, and there are supporting theoretical arguments, related to magnetic loop parameters. 
 But it is also true that different data sources 
 (e.g.\ spectroscopy and photometry) and analysis techniques (e.g.\ minimum entropy or information limit) 
 do not always lead to one clear and consistent picture (Strassmeier, 1992; Radick et al. 1998; Petit et
 al., 2004; Baliunas, 2006).

Even if real maculae are neither circular nor uniform, there will be certain 
mean values that can represent their (differential) effect to the available accuracy. 
Such mean values, as used in sunspot statistical studies, have validity in tracking and 
relating data to other activity indicators.  So while the surface structure of 
active cool stars may well be more complicated than we can presently discern, 
the approximations available can summarise observational 
findings and stimulate efforts towards more detailed future studies.  

Note that the differential maculation variation, that historically caught the attention of 
observers, should not cause the steady background component to be disregarded.  
The latter, coming from a simultaneously extant, uniform, distribution of maculae, 
can have quite a significant effect, as noted by Popper (1998) and Semeniuk (2000),
 who derived systematic differences between the distance estimates of certain cool 
 close binaries, obtained photometrically, with those from the Hipparcos satellite.  
 They found that the mean surface flux of such cool binaries was too low to allow them 
 to fit with the normal correlation from their $B - V$ colour indices and concluded  
 that a uniform distribution of dark spots could account for the difference.  
 Budding et al (2005) confirmed these results and estimated that the mean 
 surface flux could be underestimated up to a level of about 30\% in cases of 
 close binaries similar to YY Gem (see also Torres \& Ribas, 2002).
  
Computer programs that model the light curves of eclipsing variables with
surface inhomogeneities were discussed by Budding \& Zeilik (1987). This
software was developed into a user-friendly  format by M.\ Rhodes, available as
{\sc WINFITTER} from  http://home.comcast.net/$\sim$michael.rhodes/. 
The adopted technique is an iterative one that progressively defines
parameters affecting light curves, beginning with those relating to the binary
orbit, and subsequently including those controlling the extent and position of
surface spots. The procedure involves a Marquardt-Levenberg
strategy for reducing $\chi^2$ values corresponding to the given
fitting function with an assigned trial set of parameters.
If an advantageous direction for simultaneous optimization
of a group of parameters is located then that direction
can be followed in the iteration sequence (`vector search'),
otherwise the search proceeds by optimizing each parameter in turn.
For a linear problem, $\chi^2$ minimization is equivalent to the 
familiar least-squares method (cf.\ Bevington, 1969), but
the parameters in our fitting function are not
in a linear arrangement, preventing an immediate inversion
to the optimal parameter set.  However, the $\chi^2$ Hessian
is calculated numerically for a location in parameter space
corresponding to the found minimum.  If the 
location is a true minimum with all the Hessian's eigenvalues
positive, useful light on the determinacy 
of each individual parameter is thrown.

An important issue is the specification of errors.
Photometric data sets usually permit data errors to be 
assigned from the spread of differences between comparison and check
stars. We have adopted representative errors based on the observation that
the great majority of data points for YY Gem are within 20\% of the mean. 
Uniform error assignment weights the fitting at the bottom of the minima more 
highly which is beneficial in fixing the main parameters as these regions of 
the light curve have relatively high information content.

A check on 
the validity of such error estimates comes from the corresponding optimal
$\chi^2$ values.  
The ratios $\chi^2/\nu$, where $\nu$ is the number of degrees of
freedom of the fitting, can be compared with those in standard tables of the
$\chi^2$ variate (e.g.\ Pearson \& Hartley, 1954) and a confidence
level for the given model with the adopted error estimates obtained.
If $\chi^2/\nu$ is quite different from unity, we can be confident 
that either the data errors are seriously incorrect, or (more often),
the derived model is producing an inadequate representation for 
the available precision.   

This relates to another well-known aspect of optimization problems,
 i.e.\ that while a given model can be adequate to account for
 a given data-set, we cannot be sure that it is the only such model.
 This is sometimes called the `uniqueness' problem, and, in its most
 general form, is insoluble.  However, if we confine ourselves to 
 modelling with a limited set of parameters and the Hessian
 at the located $\chi^2$ minimum remains positive definite for that
 set with the $\chi^2/\nu$ ratio also within acceptable
 confidence limits, then the results are significant within the context.  
 If either of these two conditions fail
then there are reasonable grounds for doubting
the representation. Provided the conditions are met, the Hessian can be inverted to yield
the error matrix for the parameter-set.  The errors listed in Tables 2 and 3 
were estimated in this way.

 To speed up a full examination of parameter
space, the data can be binned to form normal points with phase intervals
typically 0.5\% of the period. The residuals from the  eclipse model
were first fitted with a simple two-spot model  (for procedural details see
Zeilik et al.\ 1988), but this was later revised in a fitting that included a
bright plage visible near primary minimum, on the basis of additional evidence.

The high orbital inclination of YY Gem ($\sim$86\degr) results in poor accuracy
for the spot latitude determination.  Spots of a given size at the same
longitudes but in opposite latitude hemispheres would generally show similar
light curve effects.  Attempts to derive a full spot parameter specification 
simultaneously tend to run into determinacy problems: a low-latitude spot might
be moved towards the pole in the modelling, but a quite similar pattern of
variation could then be reproduced by a corresponding decrease in size at the
same latitude. On the other hand, spot  longitudes were always fairly well
defined.

We adopted the following procedure: 
(1) Fit the eclipses for the 0.6$^m$ V light curve by adjusting the main
geometrical parameters, using the photospheric temperatures
listed by Budding \& Demircan (2007: Table 3.2).
A normalization constant also appears as a free parameter for any
given light curve. An initial value is usually adopted from 
setting the highest measured flux to a nominal value
of unity. Subsequent optimization will yield
a better representation for this.
(2) Specify initial values  for the longitudes,
latitudes and radii of spots, as in Zeilik et al.\ (1988). 
(3) Estimate the relative intensity of spots {$\kappa$} in the V band  (compared
to the unspotted photosphere).  We assigned a preliminary  value of $\kappa$
$\sim$0.2, assuming black-body emission and an approximate  mean temperature
difference of $\sim$500 K between spots and photosphere $T_{\rm p} - T_{\rm
s}$.
The low value of $\kappa$ entails that the spot size is not
so sensitive to the adopted temperature decrement for the V light curve.
Since the V spectral region lies some way to the short-wavelength side
of the Planckian peak at the adopted temperature (3770 K), only in the infra-red
will light curves start to show a noticeably decreased maculation amplitude.
This could be simulated by a smaller spot, but that would not be consistent,
of course, with radii of the same feature obtained in V.
In other words, the weight of information in the shorter wavelength photometry
goes towards fixing the spot size: at the longer wavelengths it goes towards
determining the temperature.
(4) Optimize first spot longitudes, then radii and (possibly) latitudes, using
{\sc CURVEFIT}. 
(5) Retrofit the eclipse curve for the stellar parameters with the spot
modulation removed.

\begin{table}
\begin{center}
\caption{Parameters used in or derived from the solution for the V light curve}

\begin{tabular}{lcc}
\\
\multicolumn{3}{c}{{\bf 0.6m V Light Curve model}} \\
\hline
Ratio of Luminosities & $L_{1}/L_{2}$ & 1.02$\pm$.005 \\
Ratio of Masses & $M_{1}/M_{2}$ & 1.0 \\
Ratio of Radii & $R_{1}/R_{2}$ & 1.0$\pm$.008 \\
Coeff. Limb Dark. &  $u_{1,2}$ & 0.88 \\
Radius of Primary & $R_{1}/A$ & 0.154$\pm$.001 \\
Orbital Inclination (\degr) & $i$ & 86.0  $\pm$0.11 \\
 \hline
\\
\end{tabular}
\begin{tabular}{cccc}
\multicolumn{4}{c}{{\bf Three-spot model for V light curve}} \\
\multicolumn{1}{c}{Long.} & \multicolumn{1}{c}{Lat.} & \multicolumn{1}{c}{Radius} & \multicolumn{1}{c}{Temp.\ decr.} \\
\hline
 94.8$^\circ$ & -16$^\circ$ & 16.4$^\circ$ & 0.84 \\
250.0$^\circ$ &  45$^\circ$ & 10.0$^\circ$ & 0.84 \\
342.7$^\circ$ &  21$^\circ$ & 12.3$^\circ$ & 1.13 \\
\hline
\multicolumn{2}{c}{Datum error $\Delta l$} & \multicolumn{2}{c}{0.01}\\
\multicolumn{2}{c}{Goodness of fit $\chi^2/\nu$} & \multicolumn{2}{c}{1.26} \\
\hline
\end{tabular}
\end{center}
\end{table}

Final parameters from this procedure are given in Table 2. Adopting the radial
velocity analysis of Torres \& Ribas (2002) and the standard use of Kepler's 
third law leads to a separation of the two
mass-centres as 3.898 R$_{\odot}$, or that the  radius of either star is some
0.601 R$_{\odot}$.
This is slightly less than the value Torres \& Ribas
calculated due to the difference in the two light-curve fitting results.
  Our masses 
(0.600 M$_{\odot}$), however, are in almost exact agreement with those of Torres
\& Ribas, with our own (slightly lower) value for the orbital
inclination, i.e.\ the two sets of results are within their error limits of each other.
The inclination listed in Table 2 derives from the fit to the binary light curve, however,
in the separate fitting that allows spot parameters to be estimated, a mean 
value for the inclination has been adopted. This allows the full weight of the 
difference curve data to go into the determination of the
geometrical paremeters of the starspots.The final value of $\chi^2/\nu$ given at the bottom of Table 2 
is a little high for the adopted accuracy of the data, as mentioned above. 
The photometric modelling of these V data, taken in isolation,
should then be regarded as a feasible or coarse representation of reality.
Nonetheless it is in keeping with the other results discussed in the
following sections, and the combination of evidence
gives added significance to the model.

Note that this modelling alone cannot distinguish between spots on the primary
and secondary components, particularly in the present case with an essentially
identical pair. A given spot can be situated on the primary at the
longitude indicated in  Table 2, or on the secondary at that longitude
$\pm$180\degr. The longitudes of the darkened regions are about 5 and 20 deg from
quadrature, i.e.\ they reach their maximum visibility when the two stars are not
too far from greatest elongation. This recalls Doyle \& Mathioudakis' (1990)  finding
that flares tend to occur close to quadrature phases, which, in turn, suggests a
topological connection between  flaring regions and cool photospheric spots.

\section{Models for the B, R, I and K light curves and temperatures of spots} 

\begin{table*}
\begin{center}
\caption{\bf Relative Intensities of Dark Spots in V,R,I,K and the derived temperature difference between the
spots and the photosphere}

\begin{tabular}{ccccccc}
\multicolumn{1}{c}{{\bf Filter}} &
\multicolumn{1}{c}{{\bf $\lambda_{\rm eff}$(\AA)}} &
\multicolumn{1}{c}{{\bf Limb darkening}} &
\multicolumn{1}{c}{{\bf Mean intensity}} &
\multicolumn{2}{c}{{\bf Spot Temp. Diff.($^{\circ}$K) T$_{P}$ - T$_{S}$}} & \\
& & \multicolumn{1}{c}{{\bf Coefficient}} & \multicolumn{1}{c}{{\bf $\kappa$}} &\multicolumn{1}{c}{{\bf Method 1}} &
\multicolumn{1}{c}{{\bf Method 2}} & \\
\hline
V & 5550 & 0.88 & 0.20 & \\
$R_{C}$ & 6800 & 0.73 & 0.24$\pm$.04 & 630 & 420 \\
$I_{C}$ & 8250 & 0.60 & 0.73$\pm$.04 & 200 & 280 \\
K & 22000 & 0.33 & 0.30$\pm$.08 & 1000 & 1320 \\
\end{tabular}
\end{center}
\end{table*}

Following the determination of basic parameters for the V light curve we
processed the light curves from the other filters, assuming the same geometry. We
verified that the B, R, I and K light curves could all be fitted by eclipses
having closely similar numerical values of the main parameters to those of the V. 
The large scatter of the U
band (0.6$^m$) data prevented their detailed analysis in this way. In our final
spot models for the B, R, I and K data we adopted longitudes, radii and
latitudes of the spots which were the same as for the V,  and assumed that only the
limb-darkening and mean surface brightness of the spots, relative to the
unspotted photosphere, differed. 
At a given wavelength the optimized
value of $\kappa$ corresponds to a spot mean temperature through the
implicit relation (1). The photometric information content
thus directs us towards the temperature estimate.
With limb-darkening coefficients at the mean
wavelength of the Cape/Kron R, I and Johnson B and K bands  taken from van Hamme
(1993), we determined the relative surface brightness of the spots in the
different photometric bands using {\sc CURVEFIT}. We could then estimate the
difference in temperature of the spotted regions from the unspotted 
photosphere. 
The mean surface brightness becomes adjustable in the fitting  of the infra-red
light curves. The geometrical parameters are held constant to allow the fitting to
concentrate only on the flux ratio for the infra-red data-sets. The increase in this
flux ratio can definitely be seen for the infra-red light curves, though we cannot get
away from the relatively high noise level which detracts from the temperature
estimation. The light curves are normalised in steps, with initially an approximate value
used to scale the input magnitude differences so that the out of eclipse flux level
can be approximately unity. The finally adopted fractional luminosities are then given
in terms of that corrected reference light level.

Eker (1994), discussing the determination of spot temperatures from broad-band
photometry, suggested two alternative approaches: (a) Assume that the spots and the
normal photosphere both radiate as black bodies at set temperatures.  (b) Assume
that the radiation from a spot is the same as that arising from a normal stellar
photosphere of the given  temperature, and then use the Barnes-Evans (Barnes \& Evans, 1976)
relationship between colour (V--R) and surface flux $f_V$.  Both methods can be
criticised; for example, black-body radiation is unlikely to provide a
very accurate result for localised spectral regions, considering the strong
influence of molecular bands on the flux distribution of dMe stars.   On the
other hand, spectrophotometric fluxes, predicted by current models, are not
always sufficiently close to real stellar spectra to give accurate colours  over
the temperature range required.  Given such issues, we applied the first
procedure  to the B, R, I and K light curves, and checked the result with the
second one.

In Table 3, column 4 we give the relative intensities of the spots derived from
the models fitted to the R, I and K light curves, adopting the positions and
radii of the spots from the V light curve, specifying, initially, the dark spot
intensity  as 0.2 of that of the normal photosphere. We used the following 
identity, where the left side refers to mean fluxes
$f$ in spot `$s$` and photospheric $`p`$ regions,  and the right adopts an
appropriate flux formula (e.g.\ black body) $\phi(T, \lambda)$:

\begin{center}
\begin{equation}
\frac{(f_{\lambda}/f_{V})_{s}}{(f_{\lambda}/f_V)_{p}}
= \frac{\phi(T_{\rm s},\lambda)}{\phi(T_{\rm p},\lambda)}\frac{\phi(T_{\rm p},V)}{\phi(T_{\rm s},V)}  \,\,\,  .
\end{equation}

\end{center}

\begin{figure*}
\centering
\includegraphics[width=12cm,clip,angle=0]{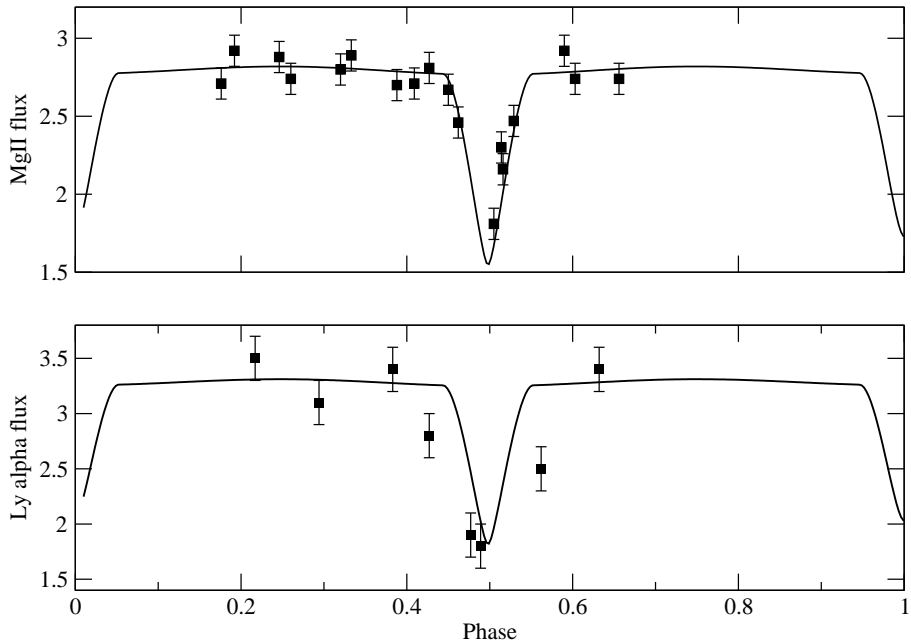}
\caption{Integrated IUE fluxes (squares) in the ultraviolet emission lines of Mg~{\sc ii} h and k (top)
 and the Ly~${\alpha}$ (bottom) 
against phase with the scaled V-band model eclipse light curves for YY Gem. The IUE fluxes are in units of 10$^{-12}$ ergs
cm$^{-2}$ s$^{-1}$. Note the reasonable fit of the Mg~{\sc ii} fluxes to the V secondary eclipse curve and the much
broader eclipse in Ly${\alpha}x
$.}

\end{figure*}

\begin{figure*}
\vspace{20cm}
  \centering
\vspace*{-16cm}

    \includegraphics[width=14cm,clip,angle=0]{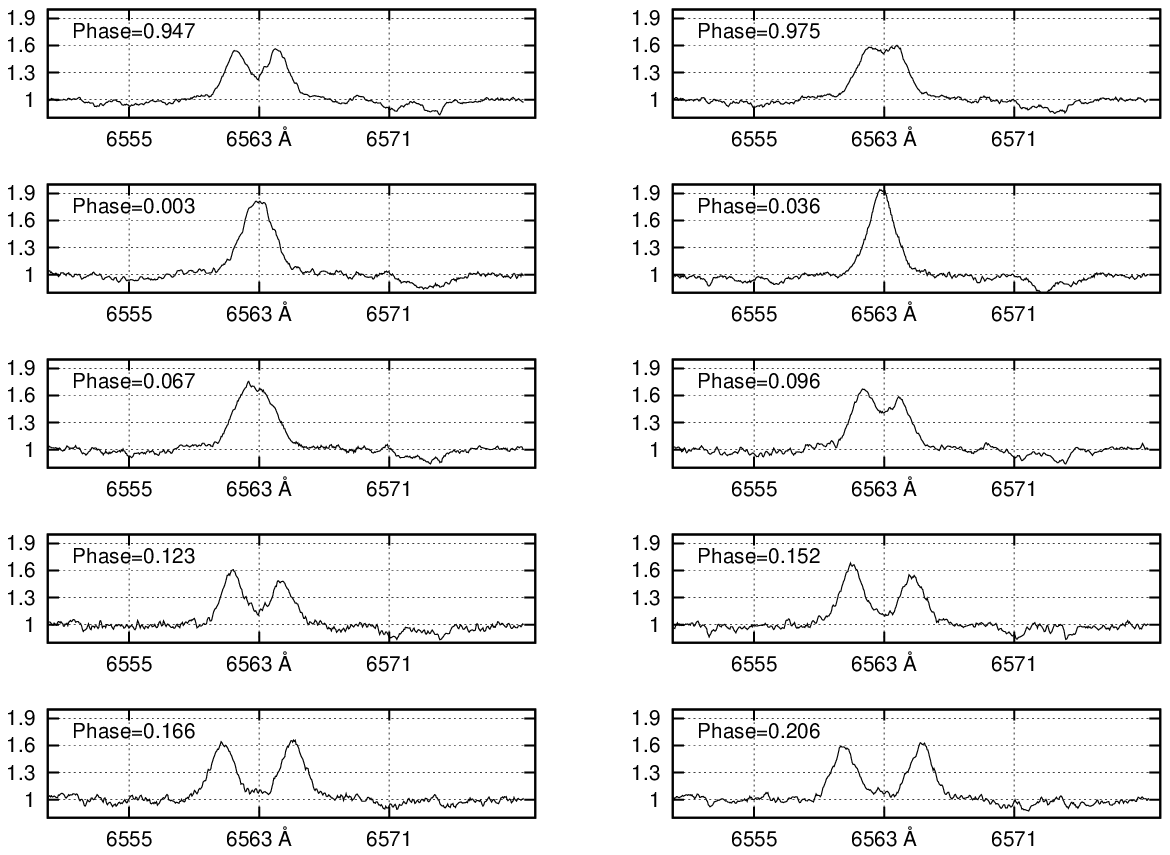}
   \includegraphics{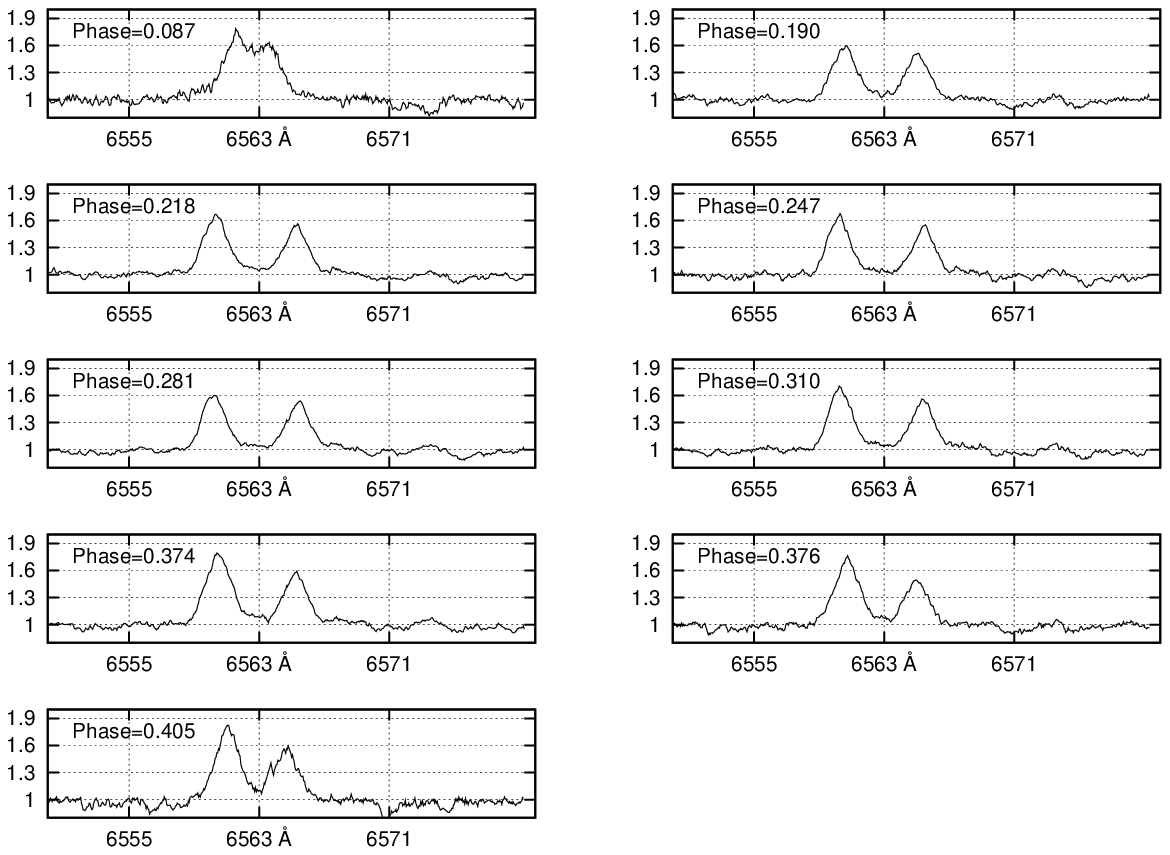}
    \includegraphics[width=14cm,clip,angle=0]{FIGURE-5B.eps}
    \vspace*{0.2cm}
 \caption{H$\alpha$ line profiles of YY Gem obtained on the 5 and 6 March 1988
 at the Crimean Astrophysical Observatory (see Tuominen et al. 1989). 
 Fluxes are normalised to the continuum.} 
\end{figure*}

Here, $\lambda$ is the effective wavelength of the R, I or K filters being 
compared with V.  The left side concerns empirically determined ratios of the
maculation amplitudes, while the  right implicitly yields corresponding spot
temperatures for given wavelengths if we have some value of the mean unspotted
photospheric temperature $T_{\rm p}$. We have adopted the value 3770 K given by
Budding \& Demircan (2007). This  was determined using an absolute flux
calibration, with an adopted flux of  $8.82\times10^{-12}$ W m$^{-2}$.  This
temperature is higher than many values for M1 stars in the literature, as
noted by Torres and Ribas (2002) who preferred a yet higher value of 3820 K.
The bolometric correction required to match the V flux is --1.18 mag.  This is
somewhat less than the value --1.25 mag that recent sources would give for
this type of star (cf.\ di Benedetto, 1998; Bessell et al., 1998), but a higher
assigned temperature would increase this discrepancy  and our adopted 3770 K
appears a reasonable compromise.  In Table 3, column 5, we list the spot temperature
differences which satisfy the aforegoing identity.

Eker's alternative approach seems less direct, given insufficient transmission
details for the three R, I and K wavebands. Here, we assume the (V--R), (V--I)
and (V--K) colours of spotted regions are the same as those of a (very cool)
star of the same spectral type or temperature.  For the R and I bands, we first
interpolated the values in Table 4  of Th\'{e} et al. (1984) to determine
spectral types of the relevant spotted regions, and thence corresponding
temperatures (Cox, 2000). For the K band we can find  a temperature directly
from the relation for log~$T_e$ to $(V-2.2\mu)$  of Veeder (1974).  Results are
given in Table 3, column 6. In both methods, the mean representative
temperatures of all dark spots affecting a given light curve are taken to be the
same.

Empirically derived values for the difference in temperature of spots and the
normal photosphere on YY Gem were found to vary from $\sim$200 K to
$\sim$1200 K, with the difference from the K band larger than that from
the R and I bands. The average results from Table 3 give a temperature difference
of 650 $\pm$ 300 K, which appears in good agreement with the  
photosphere -- spot temperature difference of 600 $\pm$ 450 K found 
by Vogt (1981) and Eker (1994) for the prototype star BY Dra (M0 V).

The disparity in the results for the mean temperatures of the spotted areas shown 
in Table 3 render infeasible an accurate resolution into distinct penumbral and umbral 
regions.  While the solar case suggests a significant role for penumbra in large spots (Bray \& Loughhead (1964)), 
the fact that derived temperature differentials for the maculation 
of cool active stars are generally less than that of large sunspots may well be an indication
that these active regions are heterogeneous in detail, either because of complex shapes and 
groupings of spots, the presence of white-light faculae, penumbral components or other, 
perhaps temporal, irregularities.

\begin{figure*}
  \centering
\vspace*{-2cm}

      \includegraphics[width=14cm,clip,angle=-90]{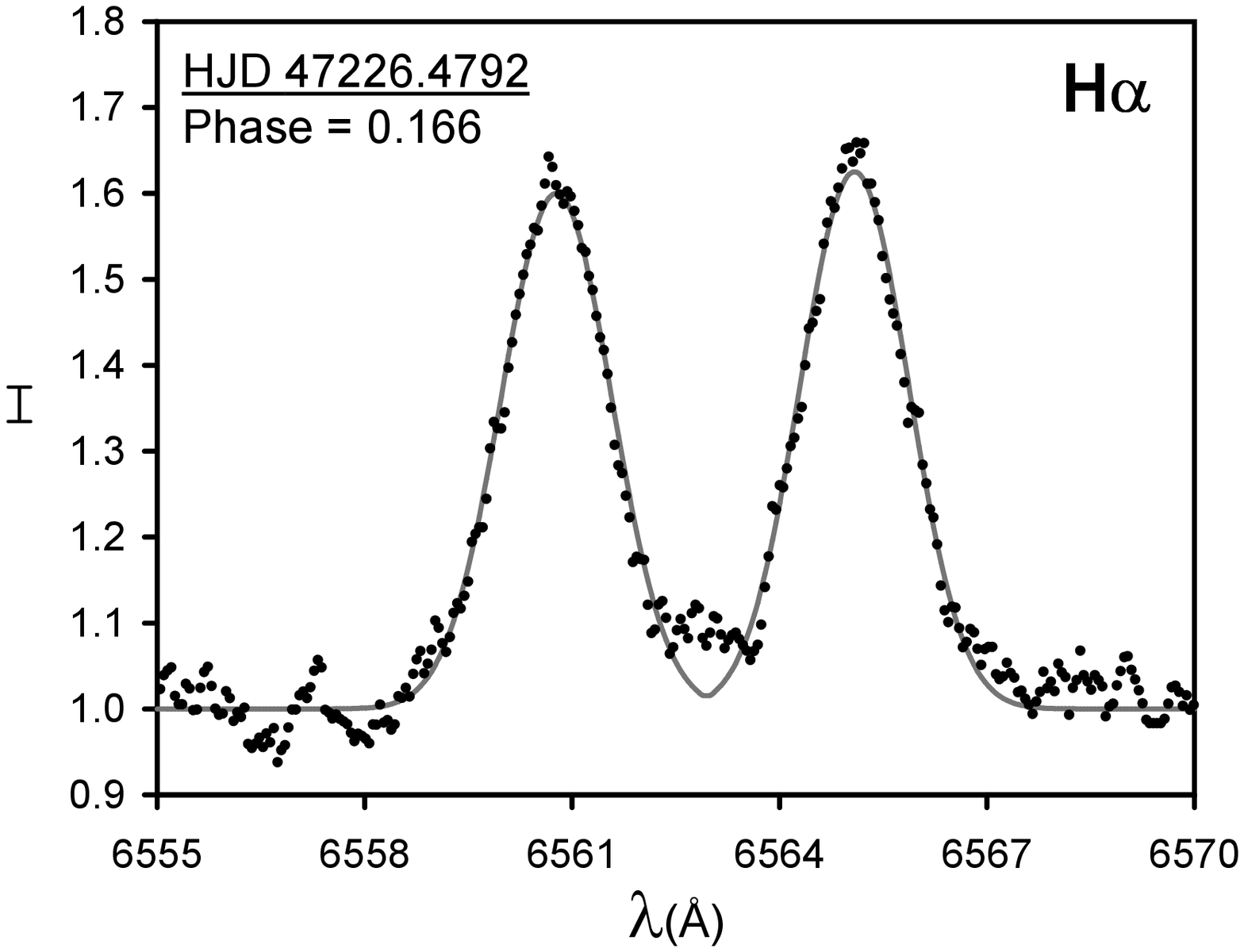}
\vspace*{-2cm}

 \caption{H$\alpha$ line profile for YY Gem with a profile model fit.} 

\end{figure*}

Several calculations of mean spot temperatures for  RS CVn stars have been
reported, giving values that differ from the normal photospheric temperature by
typically $\sim$1000 K, (Vogt, 1981; Eker, 1994, Neff et al. 1995, Olah \& Strassmeier, 2001, 
Berdyugina, 2004). 
This difference is lower  for M type  stars than for the solar type (Vogt, 1981:
Rodono, 1986). At first sight, this is not surprising, as, from their relative
areas,  one can expect penumbrae to dominate the flux. However, Dorren
(1987) argued that this is unlikely, as only spots with umbral areas $<$10\% of
the total spot area show penumbral flux domination. Dorren found, for
practically all cases, that the increased contrast  of the umbra weights the
result towards the umbral component's effect.

Berdyugina (2005) has written a comprehensive review of current techniques
for the determination of starspot temperatures. In Figure 7 of that publication, 
starspot temperature decrements (T$_P$-T$_S$) are shown to correlate with the 
photospheric temperature (T$_P$), with both dwarfs and giants scattered about 
a single mean curve. The mean starspot temperature decrement we derive here for 
YY Gem (650$\pm$300 K) falls close to the mean line in the lower part of 
Berdyugina's figure.

\section{Spectroscopic Data}

\subsection{Ultraviolet Spectra from IUE}

Observations of ultraviolet spectra from the International Ultraviolet Explorer Satellite (IUE) 
of YY Gem were scheduled for 5 and 6 March 1988 from 03:00 to 11:00 UT. A total of 30 spectra were
obtained; 9 with the short wavelength (1000-2000 A) SWP camera and 21 with the long wavelength (2000-3000 A)
LWP camera. To improve the time resolution, two exposures of 10 minutes duration were taken with the
LWP camera before the image was downloaded, whereas exposures in the SWP camera were single and longer
(circa 25m). The spectra obtained covered the secondary eclipse and some
contiguous phases. The Starlink reduction package {\sc IUEDR} was used to
extract the  emission line fluxes in Mg~{\sc ii} (2800A), Ly~${\alpha}$, 
C~{\sc iv} (1550A) and various other lines. Only the  Mg~{\sc ii} and 
Ly${\alpha}$ results will be  detailed here.

The most noticeable feature of the Mg~{\sc ii} emission during secondary 
eclipse is that it is fitted reasonably well  by the scaled V light curve (see
Figure   4 - top). This implies that the surface is approximately uniformly
covered by Mg~{\sc ii}  emitting plages. At first sight, the existence
of an approximately uniform chromosphere above an evidently non-uniform photosphere 
might be unexpected. However, this is not the first
time that such a situation has been suggested by observations. 
That the very active dMe stars may be
totally covered (`saturated') with chromospheric regions was proposed by Linsky
\& Gary (1983) as an explanation for the high integrated Mg~{\sc ii} flux on BY Dra like
stars.  Mathioudakis \& Doyle (1989) reached a similar conclusion, taking into
account the integrated Mg~{\sc ii} and soft X-ray fluxes of dM-dMe stars. 

There is some suggestion of a Mg~{\sc ii} flux increase towards the end of our IUE observation run. 
This could be associated with a  lower latitude active region at longitude about
230\degr, but there is insufficient data to confirm this suggestion.

In Figure 4 (bottom)  we show the flux in the Ly$\alpha$ line with the
geocoronal emission subtracted. The extraction was made using the method of
Byrne \& Doyle (1988). The eclipse light curve in Ly$\alpha$ appears broader
than the V-band curve.  We should note, however, that there are only a few data
points at relevant phases, and  the out-of-eclipse scatter shows deviations that
are comparable. There are two feasible explanations; one of  them intrinsic to
the star and the other not. The first alternative would  be that the broad
eclipse arises from a larger volume of Ly${\alpha}$ emitting material than the
photosphere of the  secondary star, roughly centred on that star. In effect,
this extended region would be optically thick in  Ly${\alpha}$. Another
explanation for the broad decline at secondary  minimum in Ly${\alpha}$ could
be variable absorption by interstellar H as the emission lines of the orbiting
stars are  Doppler shifted across the rest wavelength of the interstellar
absorption line.   This is unlikely to be significant, however, due to the small
range in radial velocity $\sim$15\% at eclipses when the stellar motion is
perpendicular to the line of  sight.
We conclude, therefore, that the Ly${\alpha}$ light curve reflects the
scale of the stellar outer atmosphere,  i.e.\ optically thick conditions at Ly${\alpha}$
out to heights  of the order of twice the photospheric radii (cf. Aschwanden,
2004). This is quite different to the situation for the H${\alpha}$ data that
we examine next. For H${\alpha}$, the predominant formation layer appears
to be located at a  relatively low height above the photosphere.

\subsection{Optical Spectra}

Optical spectra covering the H${\alpha}$ region were obtained with the coude
Spectrograph on the 2.6$^m$ Shajn telescope at the Crimean Astrophysical Observatory on
5 and 6 March 1988 with ten spectra obtained on the first night and nine on the
second (see Tuominen et al. 1989). The H${\alpha}$ emission at various phases
close to and out of eclipse are shown in Figures 5 and 6.  These profiles were modelled 
using the program {\sc PROF} (Olah et al, 1992) where the assumption is made
that the emission flux at a particular  wavelength is proportional to the
number of atoms in the line of sight emitting at that wavelength. The program
numerically integrates Doppler and Gaussian broadening contributions for each
of the ten slices across the  projected stellar surface.   

The results are listed in Table 4. In this table the parameters
$I_{0}$, $\lambda_{0}$,  $r$ and  $s$  give the peak intensity, mean
wavelength, rotational and Gaussian broadening coefficients respectively for 
those emission line profiles which could be separated into two components. We
note a high degree of self-consistency in the rotation parameter, implying mean
equatorial rotational velocities  of 38.6 and 38.5 km s$^{-1}$ ($\pm$0.1 km
s$^{-1}$) for the primary  and secondary respectively.  If we use the orbital
velocity sum of Torres \& Ribas (2002) and assume  co-rotation of the
components, we find a pair of essentially equal-sized stars, but with radii
some 4\% bigger  than those derived from the broadband light curves. In other
words, there is evidence for a small but  significant scale of chromospheric
enhancement from the H${\alpha}$ profiles. The scatter of the Gaussian 
component from profile to profile is significantly bigger than that of the
rotation parameter, indicating  a detectable variability of local surface
turbulence that could be associated with local inhomogeneities of velocity. 
This picture is consistent with that of the Sun where various 
H$\alpha$ emission features such as spicules and prominences etc regularly 
appear above the surface, particularly 
when active regions are close to the limb. A dMe star, such as YY Gem, with 
its much higher level of activity would be expected to show extensive 
off-disk structures

\begin{table}
\caption{Parametrisation of {\sc PROF} fittings to the
   H$\alpha$ emission lines shown in Fig 5.}
\begin{tabular}{|c|c|c|c|c|c|}
\hline
HJD & Comp.\ & $I_{0}$ & $\lambda_{0}$(\AA) & r(\AA) & s(\AA) \\ 
\hline
47226.3736 & 2 & 1.401 & 6562.848 & 0.840 & 0.854 \\ \cline{3-6}
$\phi$ = 0.036 & • & $\pm0.015$ & $\pm0.012$ & $\pm0.010$ & $\pm0.013$ \\ 
\hline
\hline
 & 1 & 0.768 & 6561.400 & 0.838 & 0.710 \\ \cline{3-6}
47226.4444 & • & $\pm0.013$ & $\pm0.017$ & $\pm0.017$ & $\pm0.021$ \\ \cline{2-6}
$\phi$ = 0.123 & 2 & 0.629 & 6564.411 & 0.838 & 0.705 \\ \cline{3-6}
 & & $\pm0.013$ & $\pm0.023$ & $\pm0.025$ & $\pm0.027$ \\
\hline
\hline
 & 1 & 0.935 & 6561.054 & 0.833 & 0.827 \\ \cline{3-6}
47226.4681 & & $\pm0.015$ & $\pm0.017$ & $\pm0.015$ & $\pm0.020$ \\ \cline{2-6}
$\phi$ = 0.152 & 2 & 0.749 & 6564.745 & 0.843 & 0.762 \\ \cline{3-6}
 & & $\pm0.013$ & $\pm0.020$ & $\pm0.019$ & $\pm0.023$ \\
\hline
\hline
 & 1 & 0.884 & 6560.765 & 0.841 & 0.789 \\ \cline{3-6}
47226.4792 & & $\pm0.014$ & $\pm0.017$ & $\pm0.015$ & $\pm0.020$ \\ \cline{2-6}
$\phi$ = 0.166 & 2 & 0.909 & 6565.089 & 0.841 & 0.777 \\ \cline{3-6}
 & & $\pm0.014$ & $\pm0.016$ & $\pm0.015$ & $\pm0.019$ \\
 \hline
 \hline
 & 1 & 0.816 & 6560.560 & 0.844 & 0.768 \\ \cline{3-6}
 47226.5125 & & $\pm0.013$ & $\pm0.018$ & $\pm0.016$ & $\pm0.020$ \\ \cline{2-6}
 $\phi$ = 0.206 & 2 & 0.837 & 6565.290 & 0.841 & 0.751 \\ \cline{3-6}
  & & $\pm0.013$ & $\pm0.017$ & $\pm0.016$ & $\pm0.020$ \\
\hline
\hline
 & 1 & 0.813 & 6560.562 & 0.847 & 0.763 \\ \cline{3-6}
47227.3132 & • & $\pm0.013$ & $\pm0.018$ & $\pm0.017$ & $\pm0.020$ \\ \cline{2-6}
$\phi$ = 0.190 & 2 & 0.682 & 6565.043 & 0.845 & 0.742 \\ \cline{3-6}
 & & $\pm0.013$ & $\pm0.021$ & $\pm0.019$ & $\pm0.024$ \\
\hline
\hline
 & 1 & 0.927 & 6560.335 & 0.844 & 0.791 \\ \cline{3-6}
47227.3361 & & $\pm0.014$ & $\pm0.016$ & $\pm0.015$ & $\pm0.019$ \\ \cline{2-6}
$\phi$ = 0.218 & 2 & 0.748 & 6565.265 & 0.830 & 0.793 \\ \cline{3-6}
 & & $\pm0.014$ & $\pm0.020$ & $\pm0.018$ & $\pm0.024$ \\
\hline
\hline
 & 1 & 0.822 & 6560.223 & 0.840 & 0.684 \\ \cline{3-6}
47227.3597 & & $\pm0.013$ & $\pm0.016$ & $\pm0.015$ & $\pm0.019$ \\ \cline{2-6}
$\phi$ = 0.247 & 2 & 0.677 & 6565.435 & 0.835 & 0.704 \\ \cline{3-6}
 & & $\pm0.014$ & $\pm0.019$ & $\pm0.019$ & $\pm0.024$ \\
 \hline
 \hline
 & 1 & 0.815 & 6560.218 & 0.843 & 0.719 \\ \cline{3-6}
 47227.3875 & & $\pm0.013$ & $\pm0.017$ & $\pm0.016$ & $\pm0.020$ \\ \cline{2-6}
$\phi$ = 0.281 & 2 & 0.705 & 6565.411 & 0.839 & 0.754 \\ \cline{3-6}
  & & $\pm0.013$ & $\pm0.020$ & $\pm0.019$ & $\pm0.024$ \\
\hline
\hline
& 1 & 0.911 & 6560.319 & 0.843 & 0.720 \\ \cline{3-6}
 47227.4111 & & $\pm0.013$ & $\pm0.015$ & $\pm0.141$ & $\pm0.017$ \\ \cline{2-6}
 $\phi$ = 0.310 & 2 & 0.720 & 6565.366 & 0.838 & 0.725 \\ \cline{3-6}
  & & $\pm0.013$ & $\pm0.019$ & $\pm0.019$ & $\pm0.023$ \\
\hline
\hline
& 1 & 1.159 & 6560.520 & 0.845 & 0.843 \\ \cline{3-6}
 47227.4410 & & $\pm0.014$ & $\pm0.014$ & $\pm0.012$ & $\pm0.015$ \\ \cline{2-6}
 $\phi$ = 0.347 & 2 & 0.770 & 6565.224 & 0.845 & 0.748 \\ \cline{3-6}
  & & $\pm0.013$ & $\pm0.018$ & $\pm0.017$ & $\pm0.021$ \\
\hline
\hline
& 1 & 1.042 & 6560.738 & 0.838 & 0.824 \\ \cline{3-6}
 47227.4646 & & $\pm0.015$ & $\pm0.015$ & $\pm0.013$ & $\pm0.017$ \\ \cline{2-6}
 $\phi$ = 0.376 & 2 & 0.661 & 6564.982 & 0.847 & 0.730 \\ \cline{3-6}
  & & $\pm0.013$ & $\pm0.021$ & $\pm0.020$ & $\pm0.025$ \\
\hline
\end{tabular} 
\end{table}

From Figure 5, we note that at quadrature (phase $\sim$ 0.25), 
when the primary component is approaching the observer (blue-shifted spectrum), the 
H${\alpha}$ line from the primary is approximately 20\% brighter than that from the secondary.
Likewise, in Table 4, we see that the ratio of the central intensities of the 
two components (primary/secondary) is around 1.2 for the three spectra near phase 
0.25 (0.218, 0.247, 0.281) rising to 1.5 for the two spectra with phase
0.347 and 0.376. 
At this phase the largest spot listed in Table 2, with longitude 94.8$^\circ$ and 
latitude $\neg$16$^\circ$ would be on the facing hemisphere of the primary component. 
A reasonable explanation for this would be that a bright H${\alpha}$ emission region on the 
primary was topologically associated with a dark spot on the same component.

\section{X-ray and Radio observations} 
\subsection{X-ray data from GINGA}

 YY Gem was observed with the Ginga (ASTRO-C) satellite, using its Large 
Area proportional Counter (LAC) 
from 09:00 UT on March 4 until 11:00 UT on March 6. Turner et al. 
(1989) give details of Ginga, and a full description of the LAC instrument including
its design, construction, calibration, operation, energy range, resolution, 
and sensitivity. 

Primarily because of Ginga's low-Earth orbit, good quality 
reduced data on YY Gem was obtained only intermittently during the overall period
listed above, in between Earth occultations, passages through the South Atlantic 
Anomaly, and times when background subtraction was otherwise relatively uncertain.
Reduction of the data was performed at Rutherford Appleton Laboratory using the 
University of Leicester's Ginga software. The time periods with best coverage and
data quality were approximately 10:00 to 15:00 UT on March 4 (see Figure 7 below), 
11:00 to 19:00 UT on March 5, and 06:00 to 11:00 UT on March 6. Sparse data were also
obtained at other intervening times, e.g. immediately after the period shown in 
Figure 7, namely from 16:00 UT on March 4 to 05:00 UT on March 5. Further details of the observations 
are given with the supplementary electronic tables.
The quiescent X-ray emission during these observations indicates two components with 
temperatures 3-4MK (soft) and 40-50MK (hard).

A flare which occurred shortly 
after the beginning of observations on 4 March had a
two-component flare  emission of 3 MK and 30-35 MK. In both the quiescent and the 
flare situations the
low-temperature (soft) component makes up  about 10\% of the total flux over
the 2-10 KeV range. Both flare and quiescent spectra showed only weak Fe~{\sc
xxv} emission at 6.7 keV.  Fits to the X-ray spectra were improved if the
Fe-abundance was reduced  to 0.5 solar. 

In Figure 7, we see that the flare begins to rise above quiescent level  
around 10:12 to 10:15 UT (HJD $\sim$ 2447224.92)  and reaches a maximum at
around 10:45 UT (HJD $\sim$ 2247224.94). The estimated integrated X-ray luminosity
is about 3-4 10$^{33}$ ergs  (2-10 KeV). The  flare light curve can be very
well represented by the magnetic reconnection model of Kopp \& Poletto (1984)
and Poletto,  Pallavicini \& Kopp (1988). For example, using a time constant of 
around 2.5
hours (i.e. the time constant of the underlying  reconnection  process) and a
start time of 10:15 UT, the predicted light curve fits the rise and peak very
well, and is a reasonable  fit to  the late decay phase. The long duration of
the X-ray event and the agreement with this model, tends to favour
interpretation as a solar-like, two-ribbon, flare event. In solar terms this 
equates to an X30-X40 class flare, equivalent to the largest solar flare ever 
observed. As with the 
increased H$\alpha$ emission mentioned in the previous section, we note that 
the flare detected in X-rays on 4 March, occurred at binary phase $\sim$0.27, 
i.e. close to quadrature and the phase of maximum visibility of the large spot 
identified in Section 4 and Table 2.

The X-ray data suggest a fairly continuous, relatively active state of the
components of YY Gem, that can probably be associated with micro-flaring, since
only a single large flare was observed during the campaign. The
background level in X-rays has a significant hard component that is even harder
than during the flares. This background may well be associated with the same
electrons that give rise to the more or less continuous microwave emission.

The flare seen in X-rays by Ginga was also observed optically with the 0.6$^m$ 
reflector on Mauna Kea. For most of the duration of the flare, observations
with this telescope were sporadic, intended for spot modulation rather than continuous
monitoring for flare activity. The U and B observations over this period are
shown as filled black squares in the lower two panels of Figure 7. With only
three or four U/B observations during the gradual phase of the flare, the precise
shape of the light curve at this time is uncertain. Nevertheless, we believe
that a reasonable estimate of the integrated optical flux during the gradual
phase of the flare can be derived from the data. We estimate the total
integrated flare energy for the gradual phase of the flare to be $8 \times 10^{33}$ ergs
and $27 \times 10^{33}$ ergs in the U and B bands, respectively. 

For a period of around 18
minutes the mode of observation was changed to continuous U-band monitoring with a time
resolution of approximately 6 seconds. This data which has been corrected for extinction and
converted to magnitudes is shown in the middle panel of Figure 7 with filled circles. By
chance, the observations captured the impulsive phase of the optical flare which sits on
top of the gradual flare described above. The integrated U-band energy of the impulsive phase
is $\sim 6 \times 10^{32}$ ergs, i.e. about an order of magnitude less than the U-band
energy of the gradual phase. Assuming the ratio of the U-band to B-band fluxes to be the same
for the impulsive stage of the flare as for the gradual phase, we derive a total optical
energy in the near ultraviolet/blue region of the spectrum for both stages of the flare of $37
\times 10^{33}$ ergs and a
ratio of X-ray to optical flux  $\sim$ 0.09. This is
close to the values for flares on UV Ceti
($\sim$ 0.2), YZ CMi ($\sim$ 0.1) and AD Leo ($\sim$ 0.1-0.2), listed by Katsova (1981). In 
her summary, Katsova (1981) notes that values L$_X$/L$_{opt}$ $\sim$ 0.1 - 0.2 are consistent 
with expectations for a flare eruption from a closed magnetic field configuration. Table 5
summarises the flare energetics from this campaign.

\begin{figure*}
\includegraphics[width=12cm,clip,angle=0]{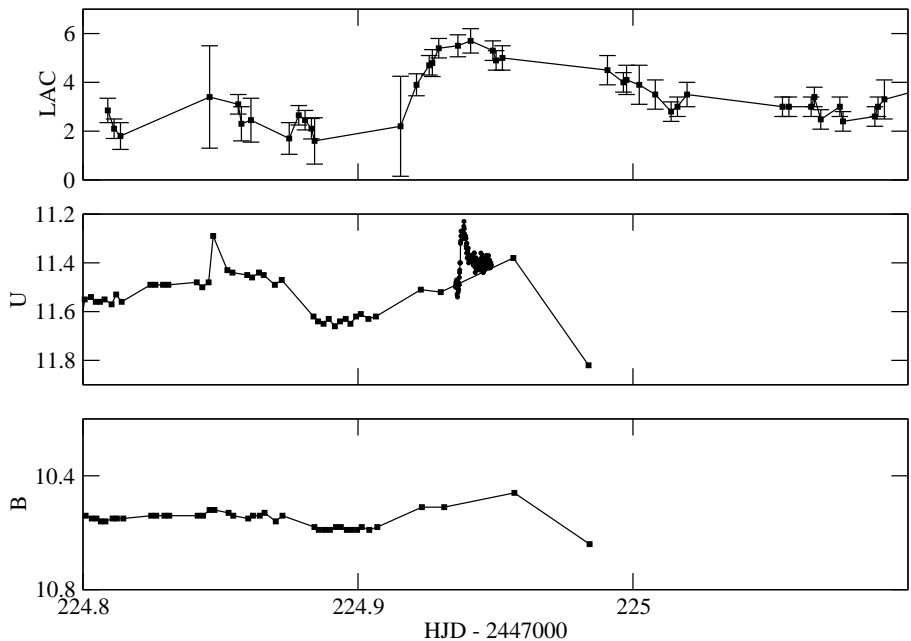}
   \caption{The flare on YY Gem observed optically and in X-rays by the 
   Ginga Satellite on 4 March 1988. Top - X-ray count rate in the Large Area Proportional Counter (LAC); Middle and Bottom - U
   and B band photometry  from 60cm telescope on Mauna Kea} 
\end{figure*}

\subsection{Microwave observations from the VLA}

YY Gem was observed in the microwave region with the VLA on two
successive days, March 4/5 and 6,  1988. The VLA was operated 
in two sub-arrays: one observing continuously at C band (6cm; 4.8 Ghz), and the other alternating
between X band (3.6 cm; 8.4 GHz) and L band (20cm; 1.46 GHz). It was detected in
both the shorter wavelengths but not at 20cm. Figure 8 shows  the 6 cm flux
curve of the system from the observations made on 4/5 March. Both  the primary
(phase = 1.0)
and secondary (phase = 1.5) eclipses appear to be visible in the flux curves with  a slight
shift of about 0.03 in the phase of minima compared to the V-band light curve.
Both radio  eclipses appear to be significantly narrower (by a factor $\sim$2) 
than the optical eclipses. The
6 cm flux at primary minimum falls  close to zero which indicates that most if
not all the 6 cm emission lies on the side of the primary  facing the secondary
or in the space between the two components. This is similar to that found on
the RSCVn binary CF Tuc by Gunn et al (1997). At secondary eclipse, (phase $\sim$1.5) 
the 6 cm flux falls to about 50\% of the normal 
non-eclipsed level. The secondary eclipse does not appear in observations made
on 6 March.  These apparent eclipses cannot be considered definitive, however,
as comparable variations are seen elsewhere in the microwave data at times
that definitely cannot be attributed to eclipses. If the feature
at phase 0.98 is an eclipse it would suggest a radio emitting region
somewhat displaced towards the advancing hemisphere.

\begin{figure*}
\includegraphics[width=12cm,clip,angle=0]{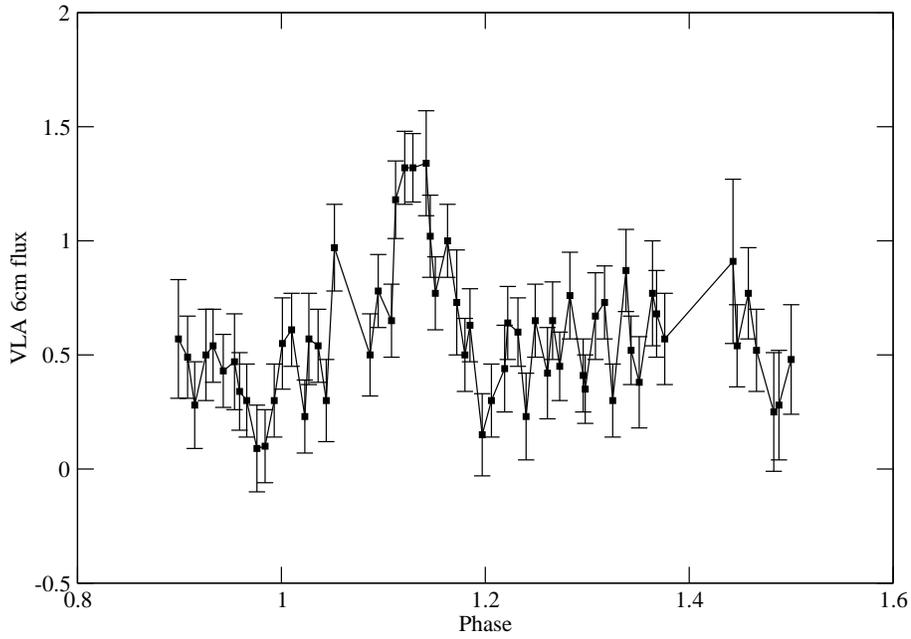}
   \caption{The 6-cm microwave flux (in mJy) from the VLA for YY Gem on 4/5 March with one sigma error
   bars} 
\end{figure*}

 On the 5th March at approximately 03:00 UT a large flare was seen in 6 cm
emission that lasted for one and a half to two hours. This 
flare was also observed in the optical at McDonald Observatory. In Figure 9
we show the 6-cm flux curve together 
with the U and B band photometry for the impulsive
part of the flare followed by its prolonged decline. We note that this flare
occurs at binary phase $\sim$0.15, i.e. not far from quadrature.

\begin{figure*}
\includegraphics[width=12cm,clip,angle=0]{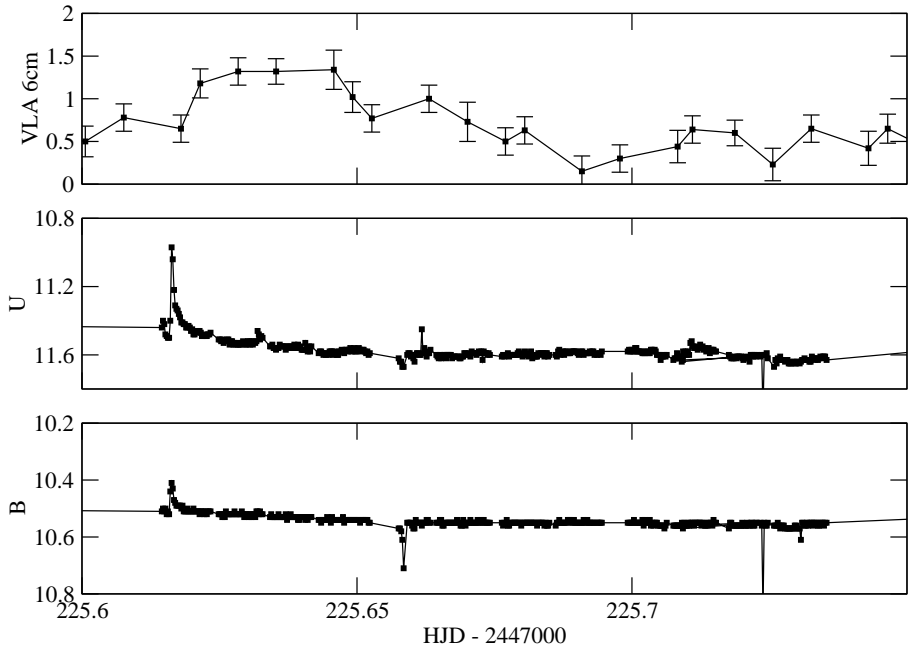}
   \caption{The flare observed in 6-cm emission on 5 March by the VLA and 
   simultaneously in the optical at the
   McDonald Observatory. Top - 6-cm microwave flux in mJy; Middle and Bottom - U
   and B band light curves} 
\end{figure*}  

From the optical light curves of the flare we derive a peak flux of
approximately 7.4 x 10$^{30}$ ergs s$^{-1}$ and 5.9 x 10$^{30}$ ergs s$^{-1}$ in U and B
respectively. Total integrated flare energies were found to be almost identical
in U and B at 2.1 x 10$^{33}$ ergs giving a total optical energy of the order 
of 5 x 10$^{33}$ ergs. One third of this energy comes from the long tail of the
flare which lasted approximately one hour in the optical and 1.7 hours in the
radio.

The peak flux at 6cm during the flare reaches 0.9 mJy, equivalent
to a total emitted flux at flare maximum of 2.4 x 10$^{14}$ ergs s$^{-1}$ Hz$^{-1}$.
The measured time-averaged radio flux, on the other hand, is 1.6 x 
10$^{14}$ ergs s$^{-1}$ Hz$^{-1}$. Using Figure 6 from Benz \& Gudel (1994), 
which relates the time-averaged X-ray and radio fluxes for
active stars, we can estimate the soft X-ray flux from the measured VLA flux 
for the above flare on YY Gem. We obtain a value for the time-averaged soft X-ray 
flux of $3 \times 10^{29}$ ergs s$^{-1}$. Assuming the flare lasted as long in X-rays 
as at 6cm radio wavelengths (namely, $\sim$ 1.7 hours), we derive an integrated 
flare energy in soft X-rays of $1.8 \times 10^{33}$ ergs. With an observed optical 
energy for the same flare of $\sim 5 \times 10^{33}$ ergs from above, we obtain a 
value of L$_{X}$/L$_{opt} \sim 0.36$,  in reasonable agreement with that 
observed directly by Ginga     described in Section 7.1 and on flares on other stars
given by Katsova (1981).

Regrettably, because the periods with successful Ginga data were so sporadic, 
 there is limited overlap between the VLA and Ginga observation windows.
 Nevertheless, it appears that Ginga did just catch the tail end of the VLA/optical 
 flare on 4 March. However, the single Ginga data point obtained at this time is
 quite inadequate to give quantitative information on the strength of the coincident 
 X-ray flare.

\begin{table*}
\begin{center}
\caption{Integrated optical and X-ray energies of flares on YY Gem in units of 10$^{33}$ ergs} 
\begin{tabular}{ccccccc}

{\bf UT (max)} & {\bf HJD (max)} & {\bf U-band} & {\bf B-band} & {\bf L$_{opt}$} & {\bf X-ray (2-10KeV)} & {\bf L$_{X}$/L$_{opt}$} \\ \hline
March 4 10:45  & 2247224.94      & 8.6          & 29           & 37              & 3-4                   & 0.09\\
March 5 03:00  & 2247225.62      & 2.1          & 2.1          & 5               & $\sim$ 1.8            & 0.36 \\
\end{tabular}
\end{center}
\end{table*}

\section{Conclusions}
 
 The March 1988 coordinated multiwavelength campaign on YY Geminorum resulted
 in less extensive ground-based coverage than originally planned. This was
 principally due to the exceptionally bad weather which prevailed in the Canary
 Islands at that time. Nevertheless, useful data were obtained from 
 ground-based facilities in the western hemisphere, notably the VLA and Mauna
 Kea in Hawaii, as well  as space-based facilities on board the IUE and Ginga
 satellites.

 With the aid of computer modelling of the V-band eclipse light curve, we
 obtain almost identical luminosities and radii for the two components of YY
 Gem and, when combined with  previously published  radial velocity curves,
 masses of $0.6M_{o}$ and radii of $0.601R_{o}$ are  derived. Fits to the
 out-of-eclipse light curves with spot models give two cool spots roughly $165^{o}$
 apart in longitude and at latitudes of $-16^{o}$ and $45^{o}$ and one 
 bright spot at $21^{o}$ latitude. Due to the high
 orbital  inclination, there is significant ambiguity in the latitudes derived
 and it is not possible to determine on which component the maculation occurs.
 Combining these models with the additional broad-band colours B, R, I and K,
 allows estimates to be made of the temperature differential between the cool spots
 and the normal photosphere. We estimate a value of $650^{o} \pm 300$, closely
 similar to previous results by Vogt (1981) on BY Draconis. Analysis of the line 
 profiles of H$\alpha$ in contemporaneous spectra of YY Gem indicates that the 
 radii of the H$\alpha$ emission volumes are
 about 4\% larger than the optical radii.

 We have observed the secondary eclipse of YY Gem in the ultraviolet emission
 lines of Mg II and Ly$\alpha$ with the IUE satellite. When scaled, the Mg II
 eclipse light curve is fitted well by  the  V-band light curve, indicating the
 presence of a chromosphere on YY Gem with uniformly emitting Mg II from
 contiguous plage regions. The existence of a widespread network of
 emission in the chromospheric lines of CaII and MgII on the Sun has been known
 for many years (see Phillips, 1992). A similar picture emerges from studies of
 many spotted stars where high levels of MgII line flux have been observed along
 with a relatively small degree of rotational modulation (see Butler et al.
 1987; Butler, 1996). The Ly$\alpha$ light curve of YY Gem, on the other hand, has a much
 broader secondary eclipse suggesting an extended outer atmosphere with a height
 of the order of twice the radius.

 YY Gem was detectable with the VLA at both 3.6 and 6 cm wavelengths but
 not at 20cm. Both the primary and secondary eclipses appear in the 6cm light
 curve on 5 March but on the following day the secondary eclipse is no longer
 evident. A poor signal-to-noise ratio prevents us from drawing any definitive
 conclusions about either eclipse at radio wavelengths, however there are
 indications that both the primary and secondary eclipses are narrower and
 slightly offset in phase in 6 cm emission compared to the optical V-band which
 would be indicative of relatively compact radio emitting regions. The almost
 total primary eclipse compared to a drop to only half the uneclipsed flux at
 the secondary eclipse would suggest that the emitting region on the primary is
 more compact than that on the secondary.

 Four flares which were detected optically on YY Gem during this campaign and 
 have been reported earlier showed evidence of periodicity
 (see Doyle et al. (1990).
 Two possible mechanisms which could have given rise to the periodicity are: (1) 
 oscillations in magnetic filaments associated with the flares, or (2) fast 
 magneto-acoustic waves between the binary components as described by Gao et al. 
 (2008).
 
 In this paper we show optical observations of two further flares on YY Gem, 
 one of which was also seen in soft X-rays by the Ginga satellite and
 the other in 6-cm radiation by the VLA. Estimates of the integrated optical 
 energy, based on the photometry are possible for both flares. For the flare 
 seen by Ginga, we can also derive the integrated soft X-ray flux, leading to
 an estimate for the ratio of the integrated X-ray to optical energy, 
 L$_X$/L$_{opt}$ $\sim$ 0.1, closely similar to the ratio observed in flares 
 on other dMe stars.
 The well established correspondence between the soft X-ray and radio fluxes 
 by Benz \& Gudel (1994) allows us to make an indirect estimate of the soft X-ray 
 energy of the flare seen by the VLA and thereby estimate L$_X$/L$_{opt}$
 for this additional flare.
 
The ratio L$_X$/L$_{opt}$ for both YY Gem flares lies close to the value 
predicted by Katsova's gas-dynamic model of a stellar flare in a constrained 
magnetic loop. If this ratio was $\sim$ 1000, as found for some solar flares, 
it would indicate, according to Katsova, an origin in an open magnetic 
structure where evaporation precluded excessive heating 
of the affected plasma.

Lastly, we note the coincidence in binary phase of several indicators of 
magnetic activity on YY Gem. The flares observed in X-rays and the radio region
occurred at phases 0.27 and 0.15 respectively; i.e. within the phase interval 
during which the largest spot at longitude 94.8$^{\circ}$ would be visible. 
Spectroscopic data obtained as part of this campaign also provides evidence 
of an H$\alpha$ emission region on the primary component with increasing 
visibility from phase 0.25 to 0.35. A reasonable interpretation of these observations
would be that all are manifestations of magnetic activity 
from a single large active region on the primary component of YY Gem. This is not 
an unexpected conclusion as similar associations between flares, H$\alpha$ emission 
regions and spots are common on the Sun and have been seen previously on other stellar 
systems (see Rodono et al. 1987, Olah et al. 1992, Butler, 1996).

\section*{Acknowledgements}

NE wishes to acknowledge support of the European Student Exchange Programme (ERASMUS) 
for a 3 month period of research and study at
Armagh Observatory. EB acknowledges stimulative input and hospitality from Armagh Observatory. We wish to thank:
the National Radio Astronomy Observatories of the USA for time on the VLA, the University 
of Hawaii for access to the 60cm telescope on Mauna Kea, the European Space Agency for observations with the 
IUE satellite, the Science and Engineering Research Council of the United Kingdom for access to UKIRT and the 
Institute of Space and Aeronautical Science of Japan (ISAS) for hospitality
and technical assistance during observations with the Japan-UK LAC instrument on Ginga. Research at 
Armagh Observatory is grant-aided by the
Department for Culture, Arts and Leisure of Northern Ireland.

\section{Appendix A - The electronic data files}

The observational data accumulated during this project and presented in the figures shown in this paper are available via the
Internet from the Armagh Observatory World-Wide Web Site using the link: http://www.arm.ac.uk/preprints/2014/654 with subdirectories
/photometry, /spectroscopy, /IUE, /VLA and /Ginga containing the optical photometry, the spectroscopic, ultraviolet, radio
and X-ray data respectively. In each subdirectory a file "notes-etc" contains a description of the data available and the
format of each file. A subdirectory /figures contains the encapsulated postscript files for the figures displayed here.

\end{document}